\documentclass[prl,aps,twocolumn,floats,superscriptaddress]{revtex4}
\usepackage{amsmath}
\usepackage{graphicx}
\usepackage{epsfig}

\newcommand{\be}{\begin{equation}}
\newcommand{\ee}{\end{equation}}
\newcommand{\bea}{\begin{eqnarray}}
\newcommand{\eea}{\end{eqnarray}}

\newcommand{\bbf}{}

\begin{document}
\title{Theory of inelastic scattering from magnetic impurities}
\author{G. Zar\'and$^1$, L. Borda$^2$, Jan von Delft$^2$, and Natan Andrei$^3$}
\affiliation{
$^1$Theoretical Physics Department, Budapest University of Technology and Economics,
Budafoki ut 8. H-1521 Hungary \\
$^2$Sektion Physik and Center for Nanoscience, LMU M\"unchen,
Theresienstrasse 37, 80333 M\"unchen, Germany \\
$^3$Center for Materials Theory, Rutgers University, Piscataway, NJ 08855, U.S.A. }
\date{\today}

\begin{abstract} 
 {\bbf We use the numerical renormalization group method} to 
  calculate the single particle matrix elements $\cal T$ of the many
 body $T$-matrix of the conduction electrons scattered by a magnetic
 impurity at $T=0$ temperature.  Since $\cal T$ determines both the
 total and the elastic, {\bbf spin-diagonal} scattering cross
 sections, we are able to compute the full energy-, spin- and magnetic
 field dependence of the inelastic scattering cross section,
 $\sigma_{\rm inel}(\omega)$. We find an almost linear frequency
 dependence of $\sigma_{\rm inel}(\omega)$ below the Kondo
 temperature, $T_K$, which crosses over to a $\sim \omega^2$ behavior
 only at extremely low energies.  Our method can be generalized to
 other quantum impurity models.
\end{abstract}

\maketitle

{\bbf Quantum mechanical phase coherence in  mesoscopic structures} 
is destroyed due to inelastic processes, where excitations such as
spin waves, electron-hole excitations, phonons, {\em etc.}, are
created in the environment with a certain probability, thus leading to
dephasing and  loss of quantum coherence after a time $\sim
\tau_\varphi$ \cite{Altshuler_review}.
%
{\bbf In some weak localization measurements of the dephasing time $\tau_\varphi$ 
down to very low temperatures, a surprising saturation}
of $\tau_\varphi$ has been observed \cite{Mohatny-Webb}.
This unexpected saturation remained a puzzle for a long time until
recently, when further experiments on mesoscopic quantum wires
confirmed that the most likely candidates to produce this surprising
behavior are magnetic impurities \cite{Birge,Glazman}. These magnetic
impurities seem to be present even in samples of extreme purity, and
unavoidably lead to inelastic scattering and the dephasing of charge
carriers.

Theoretical calculations also confirmed these expectations and showed
that the experimental data can be quantitatively explained assuming
weak inelastic scattering off Kondo impurities \cite{Altshuler,Kroha}.
These calculations, though, were performed in the weak coupling
regime, {\em i.e.}, at energies higher than the Kondo temperature,
$T_K$. However, Nozi\`eres Fermi liquid theory teaches us that well
below $T_K$ the magnetic impurity spin is screened by the conduction
electrons, and there it acts simply as a strong but conventional
potential scatterer, and thus produces no inelastic scattering.
Therefore the inelastic scattering rate from magnetic impurities is
expected to show a {\em peak} around $T_K$ and then drop to zero well
below $T_K$ \cite{JanZawa}.

These observations motivate us to study  the complete energy dependence of the inelastic scattering rate 
off a magnetic impurity. 
Here we shall focus on the simplest possible case of  
$T=0$ temperature, where the inelastic scattering rate can be defined as follows:
Assume that we have a single scattering impurity at the origin and we create an incoming flux of 
electrons with momentum ${\bf k}$, spin $\sigma$, and energy $E$ above the Fermi energy, 
far away from the origin.  This incoming flux can be scattered 
off the impurity in two different ways: ({\em i}) Either the electrons  scatter 
{\em elastically} {\bbf (both energy and spin unchanged)}
with a scattering cross section $\sigma_{\rm el}(E)$
into an outgoing  single particle state, without perturbing the environment, or
({\em ii})  they  scatter off {\em inelastically} with a corresponding cross section 
$\sigma_{\rm inel}(E)$, {\em i.e.}, and they leave  behind some electron-hole 
or spin excitations. 

In the present paper, we show how the full energy and magnetic field
dependence of $\sigma_{\rm inel}(E)$ can be determined.  The basic
idea is simple: The single particle matrix elements of the many-body
$T$-matrix, $\langle {\bf k}\;\sigma| \hat T |{\bf k}'\;\sigma'
\rangle$, determine the elastic cross section, but they are also
related to the total scattering cross section, $\sigma_{\rm tot} =
\sigma_{\rm el} + \sigma_{\rm inel}$ through the optical theorem.
Therefore, we only have to find a way to compute the $\langle {\bf
  k}\;\sigma| \hat T |{\bf k}'\;\sigma' \rangle$'s to obtain the
inelastic scattering cross section as the difference of the total and
elastic scattering cross sections:
\begin{equation}
\sigma_{\rm inel}^\sigma = \sigma_{\rm total}^\sigma - \sigma_{\rm
  el}^\sigma \; . 
\label{eq:difference}
\end{equation}
To determine $\langle {\bf k}\;\sigma| \hat T |{\bf k}'\;\sigma'
\rangle $, we shall first relate them through the reduction formulas to
some local correlation functions \cite{Itzikson}, which
we shall then calculate using the 
non-perturbative method of the numerical renormalization group (NRG)
\cite{NRG_ref}.  Though here we shall focus exclusively on the case of
$T=0$ temperature, where excitations are created from the {\bbf
  vacuum} state \cite{footnote2}, our discussions carry over, with
some modifications, to the case of finite temperatures
too \cite{follow_up}.

To be specific, we first consider the Anderson model, but our method
is rather general and applies to practically any local quantum
impurity problem. We write the Hamiltonian as $H = H_0 + H_{\rm int}$,
where $H_0$ denotes the 'free' quadratic part of the Hamiltonian,
\begin{eqnarray}
H_0 = \sum_{\sigma}
\epsilon_d  d^\dagger_\sigma d_\sigma 
+ \sum_{\sigma,\bf k} 
\xi({\bf k}) c^\dagger_{{\bf k}\;\sigma}c_{{\bf k}\;\sigma} 
\;,
\nonumber
\end{eqnarray}
and $H_{\rm int}$ stands for the on-site Hubbard interaction and hybridization
\begin{equation}
H_{\rm int} = U \; n_\uparrow n_\downarrow
+   V  \sum_{\sigma,\bf k} 
\bigl( c^\dagger_{{\bf k}\;\sigma} d_\sigma + {\rm h.c.}
\bigr)
\;, 
\end{equation}
with $n_\sigma = d^\dagger_\sigma d_\sigma$. The operator
$c^\dagger_{{\bf k}\;\sigma}$ creates an electron in a plane wave
state with momentum ${\bf k}$, energy $\xi({\bf k})= \frac{{\bf
    k}^2}{2m} - \mu$ and spin $\sigma$, while $d_\sigma$ is the
annihilation operator of the $d$-electron.

We proceed to define incoming and outgoing {\it scattering states} as
well as the corresponding field operators and Hilbert
spaces \cite{Itzikson}. As the impurity is local, the interaction
switches off far away and the 'in' and 'out' states are eigenstates of
the full Hamiltonian with the asymptotic boundary condition of behaving as
 plane waves in the $x \to -\infty$ and $ x \to \infty$
limits, respectively.   The
many-body $S$-matrix and the $T$-matrix elements are then simply defined in
terms of the overlaps of the incoming and outgoing scattering states,
\begin{eqnarray}
&&\langle b, {\rm out} | {}a, {\rm in}\rangle  \equiv 
\langle b, {\rm in}|\hat S| {}a, {\rm in}\rangle\;,\\
&&\hat S  =  1 + i\; \hat T\;.
\end{eqnarray}
 In the interaction representation, 
the explicit form of the $S$-matrix is given by the well-known expression 
$\hat S =  {\rm T}\exp \{-i \int_{-\infty}^\infty  H_{\rm int} (t) \;dt\}$,
where ${\rm T}$ is the time ordering operator.


\begin{figure}[htb]
\begin{center}
\epsfxsize8cm
\epsfbox{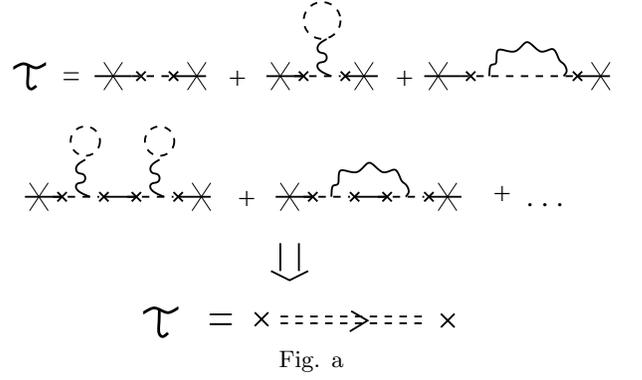}
Fig. a
\vskip0.5cm
\epsfxsize8cm
\epsfbox{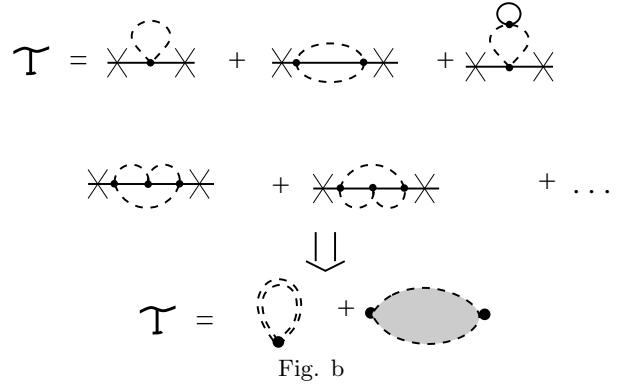}
Fig. b 
\end{center}
\vskip0.1cm \caption{\label{fig:diagrams} (a) Diagrammatic derivation
  of Eq.~(\ref{eq:anderson_tmatrix}). Dashed and continuous lines
  denote the bare propagators of the $d$-level and the conduction
  electrons, {\bbf small fat} crosses stand for hybridization $V$, and
  wavy lines denote the on-site interaction $U$. (b) Diagrammatic
  representation of the $T$-matrix in the Kondo problem. Dashed lines
  denote pseudofermion propagators and describe the evolution of the
  impurity spin, while continuous lines denote free conduction
  electron propagators.  Filled circles stand for the exchange
  interaction $J$.  The first term of the $T$-matrix is simply
  proportional to the expectation value of the impurity spin, it is
  frequency independent, and vanishes in the absence of magnetic
  field.  The second term can be identified as the composite fermions
  correlation function.}
\end{figure}

Since we are primarily interested in the single-particle matrix
elements of $\hat T$, we consider the case where asymptotically $|{}a,
{\rm in}\rangle$ is simply a single particle low energy excitation of
the free {\bbf vacuum} defined by $H(x=-\infty) \to H_0$, having
momentum ${\bf k}$ and spin $\sigma$:
$
|{}a, {\rm in}\rangle = | {\bf k},\sigma \rangle \;$.
Then the single-particle matrix elements of the $T$-matrix read
\begin{equation}
\langle {\bf k},\sigma| \hat T | {\bf k}',\sigma' \rangle 
= 2\pi \delta(\xi({\bf k}) - \xi({\bf k}')) 
 \langle {\bf k},\sigma|{\cal  T} | {\bf k}',\sigma' \rangle \;,
\end{equation}
where we separated a Dirac delta contribution due to energy
conservation and defined the on-shell $T$-matrix $\cal T$. Next,
following the manipulations of Ref.~\onlinecite{Itzikson} we
re-express the off-diagonal (${\bf k} \ne {\bf k}'$) matrix elements
of the $T$-matrix as:
\begin{eqnarray}
\label{eq:T-matrix}
\lefteqn{\langle {\bf k}\;\sigma| {\cal T} | {\bf k}' \sigma'\rangle =
  } \\
& & - s\; G_0^{-1}(\xi, s\;{\bf k})\; G_{s\sigma, s \sigma'}
(\xi, s\; {\bf k}, s \;{\bf k}') \; G_0^{-1}(\xi, s\;{\bf k}')\;,
\nonumber
\end{eqnarray}
where $s={\rm sgn}\; \xi$ distinguishes electron-like excitations from
hole-like excitations, $G_0^{-1} = i \frac{\partial}{\partial t} + \mu
+ \frac1{2m} \nabla^2$ denotes the inverse of the free Green's
function, and $G$ is the time-ordered single particle Green's
function.  The meaning of Eq.~(\ref{eq:T-matrix}) becomes more
transparent in the diagrammatic language of Fig.~\ref{fig:diagrams}:
{\bbf As indicated by the large, thin crosses there, one has to drop}
the contributions of the two external legs of all
scattering diagrams to the single electron Green's function and the
rest is just the on-shell single particle matrix element of the
many-body $T$-matrix.  In the particular case of the Anderson model,
$\cal T$ does not depend on the direction of incoming and outgoing
momenta, and a simple Dyson equation relates it to the $d$-level's
time-ordered propagator (see Fig.~\ref{fig:diagrams})
\begin{equation}
{\cal T}^\sigma(\omega)  = - s\; V^2 \; G_{d}^{s \cdot \sigma} (\omega)\;,
\label{eq:anderson_tmatrix}
\end{equation}
where $s={\rm sgn}\, \omega$ and we allowed for spin-dependence due to
the presence of an external magnetic field $B$. 
Eq.~(\ref {eq:anderson_tmatrix})
has also been derived in a somewhat different way in
Ref.~\onlinecite{schrieffer}.

{\bbf According to the optical theorem, the spin-dependent
total scattering cross section is given by the
imaginary part of the diagonal matrix elements of the ${\cal T}$-matrix:}
\begin{equation}
\sigma_{\rm total}^\sigma = \frac{2}{v_F} {\rm Im} \langle {\bf p} \sigma | {\cal T} | {\bf p} \sigma\rangle
\;,
\end{equation}
{\bbf where }
$v_F$ denotes the Fermi velocity.
The elastic scattering cross section, on the other hand, is related to
the square of $\cal T$:
\begin{equation}
\sigma_{\rm el}^\sigma = \frac{1}{v_F} \int {d {\bf p}' \over (2\pi)^3}
2\pi \; \delta(\xi'-\xi) |\langle {\bf p}' \sigma | {\cal T} | {\bf p}
\sigma\rangle|^2 \;. 
\end{equation}
Once these two cross-sections are known, we can compute the inelastic
cross section $\sigma_{\rm inel}$ through Eq.~(\ref{eq:difference}).

It is instructive to rewrite $\sigma_{\rm inel}$ in case of the
Anderson model. For electrons we have
\begin{equation}
\sigma_{\rm inel}^{\sigma}(\omega>0) = {4\pi\over k_F^2} \Bigl[
-{\Gamma\over 2} {\rm Im} \;G_d^\sigma -  \Bigl({\Gamma\over 2}\Bigr)^2 |G_d^\sigma|^2\Bigr]\;,
\end{equation}
where $\Gamma = 2\pi V^2 \varrho_0$ denotes the width of the $d$-level
and $\varrho_0 = k_F^2/ 2 \pi^2 v_F$ is the conduction electrons'
density of states for one spin direction.
For $B=0$, this expression reduces to 
\begin{equation}
\sigma_{\rm inel}(\omega>0) = {2\pi\over k_F^2}  {\Gamma\;(-  \Sigma''(\omega))
\over (\omega - \epsilon_d - \Sigma'(\omega))^2 + (\Sigma''(\omega) - {\Gamma\over2})^2}\;,
\nonumber
\end{equation}
where $\Sigma'$ and $\Sigma ''$ denote the real and imaginary parts of
the $d$-propagator's self-energy. The analytical properties of the
Green's function imply that the above expression is always positive
and it only vanishes where $\Sigma ''$ becomes zero. Furthermore, the
Fermi liquid theory of Yamada and Yoshida tells us that $\Sigma'' \sim
\omega^2$ {\bbf as $\omega \to 0$,} and thus the inelastic scattering rate
vanishes as $\omega^2$ at the Fermi energy. Note that at the same time
the total scattering cross section approaches the unitary limit.

To compute the full behavior of $\sigma_{\rm inel}$ we determined the
${\cal T}$-matrix using the numerical renormalization group method
\cite{NRG_ref}.  Since we were dominantly interested in the low-energy
universal regime of the Anderson model, we took the
$U/2=-\epsilon_d\to \infty$ limit and performed the calculations using
the Kondo Hamiltonian,
\begin{equation}
H_K = {J\over 2}  \sum_{{\bf k}, {\bf k}'}
{\vec S} \cdot (c^\dagger_{{\bf k} \sigma }{\vec \sigma}_{\sigma
  \sigma'} c^\dagger_{{\bf k'} \sigma'})\;.
\end{equation}
Up to an overall normalization factor, the low-energy part of the
spectral function of the original $d$-level propagator in the Anderson
model can be shown to correspond to the spectral function of the
following composite fermion operator in the Kondo model \cite{Theo},
$F_\sigma \equiv \sum_{\sigma', {\bf k}} {\vec S} \cdot {\vec
  \sigma}_{\sigma\sigma'} c_{{\bf k} \sigma'}$.  (For a diagrammatic
proof, see Fig.~\ref{fig:diagrams}).  The imaginary part of $G_d\sim
\cal T$ can thus be determined by simply computing the spectral
function $\varrho_F(\omega)$ of the composite fermion numerically, and
then a Hilbert transform can be used to get the real part of $G_d$ and
thus the full $T$-matrix.  In all these calculations it is essential
to have high quality data.  It is also crucial to determine the
normalization factor of $\Gamma G_d$ correctly. This can be done by
using the Fermi liquid relation, $-{\rm Im} \; \Gamma
G^\sigma_d(\omega=0^+) = 2 \sin^2 \delta_\sigma$, with $\delta_\sigma$
the phase shift at the Fermi energy. We extracted the latter directly
from the finite size NRG spectrum of the Kondo model
\cite{NRG_ref,Hofstetter}.

\begin{figure}[tb]
\begin{center}
\epsfxsize8cm
\epsfbox{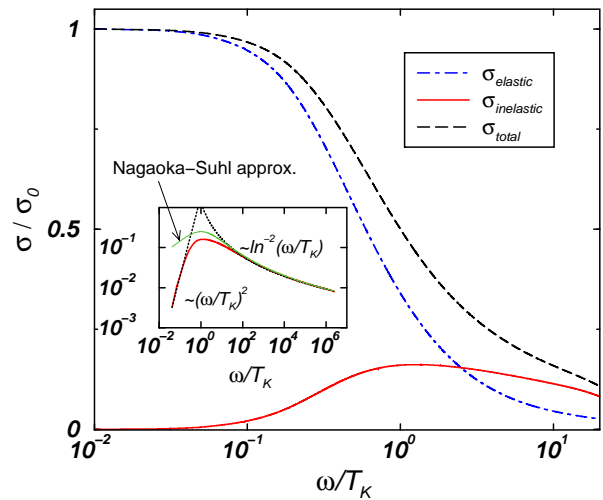}
\end{center}
\vskip0.1cm \caption{\label{fig:inelastic} Inelastic, elastic, and
  total scattering cross sections in units of $\sigma_0 = 4\pi/k_F^2$
  at $T=0$ and in the absence of magnetic field,
as a function of the logarithm of the energy of the
  incoming electron.   Only the
  electronic contribution ($\omega>0$) is plotted. $\sigma_{\rm inel}$
  {\bbf only has a very weak (logarithmic) energy dependence} above
  $T_K$, scales approximately linearly with $\omega$ for $0.01\;T_K <
  \omega < T_K$, and scales as $\sim \omega^2$ for $\omega<0.01\;T_K$.
  The inset shows the $\sim \omega^2$ and $\sigma_{\rm inel}\sim
  1/{\rm ln}^2(T_K/\omega)$ regimes for $\omega \ll T_K$ and $\omega
  \gg T_K$, respectively.
}
\end{figure}

\begin{figure}[tb]
\begin{center}
\epsfxsize8cm
\epsfbox{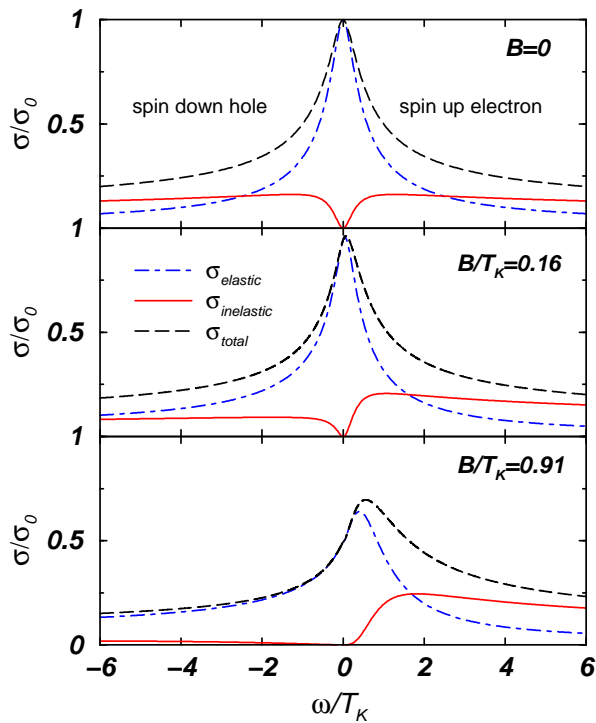}
\end{center}
\vskip0.1cm \caption{\label{fig:inel_magnetic} Energy dependence of
  spin-dependent elastic and inelastic scattering rates in units of
  $\sigma_0 = 4\pi/k_F^2$, at $T=0$ and in the presence of a local magnetic
  field $B$.
}
\end{figure}

Our results for the case of no external magnetic field are shown in
Fig.~\ref{fig:inelastic}.  Most of the scattering is inelastic at
energies above the Kondo energy, $|\omega| > T_K$. Decreasing the
energy of the incoming electrons (holes), the total scattering cross
section increases and, at energies below the Kondo scale, it finally
saturates at a value $\sigma_0 = 4\pi/k_F^2$.  This behavior must be
contrasted to the inelastic scattering rate, which slowly increases as
$\omega$ decreases, has a broad maximum around $T_K$ then suddenly
drops and vanishes at the Fermi energy.  On linear energy scales (see
Fig.~\ref{fig:inel_magnetic}), $\sigma_{\rm inel}$ varies rather
slowly above $T_K$, is very large even at $\omega \sim 20 T_K$, and
vanishes {\bbf rather} suddenly around $\omega\sim T_K$.  For very
small energies $\sigma_{\rm inel} \sim \omega^2$, in agreement with
Fermi liquid theory, however, this quadratic behavior appears only at
very low energies, and $\sigma_{\rm inel}$ is almost linear for $0.1
\;T_K <\omega < T_K$.  At energies $\omega \gg T_K$ the inelastic rate
is simply dominated by {\bbf energy-conserving} spin-flip scattering,
and is therefore expected to scale as $\sim 1/{\rm ln}^2(T_K/\omega)$,
as we indeed find numerically. Note that the Nagaoka-Suhl 
approximation~\cite{Zawareview}  
is only appropriate for $\omega \gg T_K$ (see inset of Fig.~2).

We also computed the inelastic scattering rate in the presence of a
local magnetic field, $B$, directed downwards along the $z$-axis (see
Fig.~\ref{fig:inel_magnetic}).  In this case there is a dramatic
difference between the inelastic scattering properties of spin up and
spin down particles. Already a small field, $B \sim 0.1 T_K$ results
in a strong spin-dependence of the inelastic scattering, but for
$B\sim T_K$, this difference is even more dramatic. At this field the
spin of the magnetic impurity is practically aligned with the
external field and points downwards.  Therefore an incoming spin down
particle (electron or hole) is unable to flip the impurity spin.  More
precisely, only 
{\bbf higher order} inelastic processes can result in a
flip of the local impurity spin.  This is, however, not true for spin
up particles: An incoming spin up electron can exchange its spin with
the magnetic impurity in a first order process, 
resulting in a maximum in the
inelastic scattering cross section around $\omega \sim B$ for spin up
electrons and holes and a very broad inelastic background for $\omega
> B$.

Though here we mostly focused on the simplest cases of
the Anderson  and the single channel spin $S=1/2$ Kondo models at
$T=0$ temperature, our formalism can be extended to other models and
to finite temperatures as well \cite{follow_up}.  In particular, while
for some quantum impurity models no simple diagrammatic theory is
available, the composite fermion's spectral function can be computed
in any Kondo-type model to obtain the matrix elements of the
$T$-matrix, and the renormalization group flow of the eigenvalues of
the $S$-matrix can be studied in all these cases \cite{follow_up}.
While usually a thorough numerical analysis is needed to understand
the full behavior of $\sigma_{\rm inel}$, in some models
simple analytical results can also be obtained. In the specific case
of the 2-channel Kondo problem, {\em e.g.}, we know that the single
particle matrix elements of the $S$-matrix identically vanish at the
Fermi energy, $\omega=0$ \cite{Maldacena,Jan}.  This implies that
$\varrho_0 {\cal T}^{\rm 2CK}(\omega=0) = i/\pi $ and leads to the
rather surprising relation at the Fermi level,
$
\sigma_{\rm inel}^{\rm 2CK}=\sigma_{\rm el}^{\rm 2CK}=
{\sigma_{\rm tot}^{\rm 2CK}/ 2}
$:
Though the $S$-matrix vanishes identically,
half of the scattering processes remain elastic. The non-vanishing
inelastic scattering rate is characteristic of non-Fermi liquid
quantum impurity models.  Application of any finite magnetic field
drives the 2-channel Kondo model to a Fermi liquid fixed point, and
gives rise to a vanishing inelastic scattering rate at the Fermi
energy.

We have to emphasize that, though they must be related, the inelastic
scattering rate we computed is {\em not} identical to the dephasing
rate measured in weak localization experiments \cite{Mohatny-Webb},
since {\bbf the former} incorporates spin-flip scattering processes as
well as the creation of electron-hole pairs. While we only computed
$\sigma_{\rm inel}(\omega, T=0)$, we expect that $\sigma_{\rm
  inel}(\omega=0, T)$ has a very similar form. In this sense, our
finding that the inelastic scattering rate is roughly linear with
$\omega$ for $0.05 \;T_K < \omega < 0.5 T_K$ agrees 
{\bbf qualitatively} with the recent
experimental results of Ref.~\onlinecite{bauerle}.

In summary, we have shown how the full energy and magnetic field
behavior of the $T=0$ inelastic scattering rate can be computed by
exploiting the reduction formulas and then using the powerful
machinery of numerical renormalization group to compute the single
particle matrix elements of the many-body $T$-matrix. We have shown
that the Fermi liquid theory of Yamada and Yoshida directly implies a
quadratically vanishing inelastic scattering rate at the Fermi energy
in the specific case of the Anderson model. Scattering properties of
the Kondo model have been computed by calculating the composite
Fermion's spectral function.  Our numerical calculations show that the
abovementioned $\sigma_{\rm inel} \sim \omega^2$ regime appears only at
energies well below the Kondo temperature. In a magnetic field $B>T_K$
the inelastic scattering is very sensitive to the spin of the
scattered single-particle excitation.

We are grateful to L. Glazman, L. Ioffe, A. Jakov\'ac, and A.
Zawadowski for valuable discussions.  This research has been supported
by NSF-MTA-OTKA Grant No. INT-0130446, Hungarian Grants No. OTKA
T038162, T046267, and T046303, and the European 'Spintronics' RTN
HPRN-CT-2002-00302.  G.Z. has been supported by the Bolyai Foundation.


\begin{references}
\bibitem{footnote2} 
For a degenerate  interacting {\bbf vacuum} state, we have to trace over 
the various {\bbf vacuum} states in the following discussions.
\end{references}
\end{document}